\newcommand{\beq} {\begin{eqnarray}}
\newcommand{\eeq} {\end{eqnarray}}
\documentclass[12pt, preprint,preprintnumbers,amsfonts,aps,nofootinbib]{revtex4}
\usepackage{graphicx}
\begin{document}

\title{Sequestering the Gravitino: Neutralino Dark Matter in Gauge Mediation}

\author{Nathaniel J. Craig}
\email{ncraig@stanford.edu}
\affiliation{Department of Physics, Stanford University, Stanford, CA 94305-4060} 

\author{Daniel Green}
\email{drgreen@stanford.edu}
\affiliation{SLAC and Department of Physics, Stanford University, Stanford, CA 94305-4060}

\preprint{SU-ITP-08/17}
\preprint{SLAC-PUB-13363}

\begin{abstract}

In conventional models of gauge-mediated supersymmetry breaking, the lightest supersymmetric particle (LSP) is invariably the gravitino. However, if the supersymmetry breaking sector is strongly coupled, conformal sequestering may raise the mass of the gravitino relative to the remaining soft supersymmetry-breaking masses. In this letter, we demonstrate that such conformal dynamics in gauge-mediated theories may give rise to satisfactory neutralino dark matter while simultaneously solving the flavor and $\mu / B \mu$ problems.

\end{abstract}

\maketitle
\section{Introduction}

In supersymmetric model building, there is often a tension between the experimental constraints of flavor physics and dark matter.  A cold relic with the observed relic abundance of $\Omega_{DM} \simeq 0.25$ \cite{Dunkley:2008ie, Tegmark:2006az} naturally suggests a cross section consistent with a weakly interacting particle of mass $m \simeq 100-1000$ GeV.  For such a particle to be a good dark matter candidate, it should be stable.  It is often assumed that this can be accomplished in supersymmetry (SUSY) by imposing R-parity.

However, it is often the case that solving the flavor problem in the MSSM implies a gravitino LSP. The gravitino mass arises from gravity-mediated effects, which in general are flavor violating. Experimental constraints on flavor violation require that gravity-mediated contributions to the soft SUSY-breaking masses are small; if additional physics is responsible for generating the remaining soft masses, one might expect the gravitino to be much lighter than the other superpartners.

From this point of view, the most natural means of communicating SUSY-breaking to the MSSM would seem to be gauge mediation \cite{Dine:1981za,Dimopoulos:1981au, Nappi:1982hm,AlvarezGaume:1981wy,Dine:1995ag}.  In gauge mediation, the MSSM soft masses are generated through gauge interactions and are therefore flavor blind.  One can then easily arrange for the scale of supersymmetry breaking to be low enough to render gravity-mediated contributions negligible. Consequently, the gravitino is generally quite light, and thus it is often stated that a gravitino LSP is one of the firm predictions of gauge mediation \cite{Giudice:1998bp, Meade:2008wd}. Unfortunately, gravitino dark matter is somewhat untenable, both in terms of cosmology and direct detection.  There has been much ongoing work to engineer satisfactory dark matter candidates within gauge mediation \cite{Dimopoulos:1996gy, Nomura:2001ub, Fujii:2002fv, Ibe:2006rc,Feng:2008zza, Feng:2008ya,Dudas:2008eq}.  With many direct detection searches looking for a massive particle recoil \cite{Akerib:2005kh, Angle:2007uj}, it is unfortunate that most of these models do not predict this classic observable signature.

In models with strongly coupled hidden sectors, the na\"{i}ve scale of the gravity mediated contribution to soft masses is known to be inaccurate.  Indeed, in models of anomaly mediation, it is necessary to suppress gravity-mediated contributions when SUSY breaking occurs at a high scale \cite{Randall:1998uk, Giudice:1998xp}.  Such suppression is generally accomplished by ``sequestering.'' In models with extra dimensions, sequestering is obtained by separating the SUSY breaking from the Standard Model by a large distance.  There is also a purely four-dimensional picture known as conformal sequestering \cite{Luty:2001jh,Luty:2001zv}, in which gravity-mediated operators are suppressed by large anomalous dimensions.

For the most part, sequestering has been considered strictly in the context anomaly mediation.  However, it was recently observed \cite{Roy:2007nz,Murayama:2007ge} that by incorporating sequestering in models with gauge mediation, the $\mu / B\mu$ problem may be solved.

In this letter, we will discuss the implications of sequestered gauge mediation for dark matter.  In particular, we will show that the combination of sequestering and gauge mediation can simultaneously solve the flavor problem and the $\mu / B\mu$ problem while furnishing a neutralino dark matter candidate with the right relic abundance.   

The organization of this paper is as follows: In Sec. \ref{sec:seq} we review sequestering and its application to gauge mediated supersymmetry breaking, while in Sec. \ref{sec:theories} we consider a general class of conformally-sequestered theories suitable to solving $\mu/ B \mu$ in which the gravitino is no longer the LSP. Such theories lead to a characteristic spectrum of soft masses at the scale of conformal symmetry-breaking.  Finally, in Sec. \ref{sec:dm}, we discuss the resultant weak-scale sparticle spectrum and explore the range of parameters for viable neutralino dark matter. 

\section{Sequestered Gauge Mediation \label{sec:seq}}

In both gravity and gauge meditation, the objects of interest are operators that mix the Standard model and SUSY-breaking fields. Such operators give rise to supersymmetry-breaking soft masses in the MSSM, and obtain the general form
\beq
\label{operators}
c_{\phi} \int d^4 \theta \frac{S^{\dag} S}{M^2} \phi^{\dag} \phi \qquad  \qquad c_{W} \int d^2 \theta \frac{S}{M} W_{\alpha} W^{\alpha}
\eeq
where $S$ is a gauge singlet that develops a SUSY-breaking F-term. Here $\phi$ is an MSSM chiral superfield representing any squark or slepton, while $W_{\alpha}$ are superfields of the MSSM gauge multiplets.  By integrating out the supersymmetry-breaking $F$-term $F_{S}$, these operators generate SUSY-breaking soft masses for the squarks, sleptons and gauginos.  In gravity mediation, the mediation scale is taken to be $M = M_{Pl},$ and there is no reason to suppose the coefficients $c_{\phi}$ respect the flavor symmetries of the MSSM. Such flavor-violating terms are quite dangerous, and frequently give rise to flavor-changing neutral currents (FCNCs) in violation of experimental bounds.  In the case of gauge mediation, however, the scale $M$ is the messenger mass and the coefficients of (\ref{operators}) {\it are} flavor blind due to the family invariance of gauge interactions. Such flavor-blindness is among the most compelling features of gauge-mediated supersymmetry breaking. 

{\it A priori,} one would typically expect the coefficients of (\ref{operators}) to be $\mathcal{O}(1).$ However, sequestering in the hidden sector may lead to significant suppression of the coefficients $c_\phi, c_W.$ Broadly speaking, sequestering can be characterized as a dynamical mechanism that forces \emph{all} the coefficients $c \ll 1$. Such sequestration has been shown to arise in the context of extra dimensions \cite{Randall:1998uk} and strongly coupled field theories \cite{Luty:2001jh,Luty:2001zv}.  In higher-dimensional theories, the suppression of mixing coefficients occurs by separating the SUSY breaking from the MSSM by large distances. In such theories, the physical separation leads to small couplings between spatially localized fields, although this can be difficult to obtain in unwarped models \cite{Kachru:2007xp}. 

However, these extra-dimensional constructions are often analogous to a strictly four-dimensional picture, in which the suppression of coefficients is due to large wavefunction renormalization from strictly four-dimensional dynamics. Here we will focus on the purely four dimensional version known as conformal sequestering \cite{Luty:2001jh,Luty:2001zv}, which can be explicitly dual to an extra-dimensional description where the effects are calculable \cite{Kachru:2007xp}.  Conformal sequestering may arise in theories where the SUSY breaking sector is strongly coupled.  In these cases, the operators $S$ and $S^{\dag} S$ may acquire large anomalous dimensions.\footnote{Strictly speaking, $S$ is a chiral operator, so its conformal dimension is protected.  However, at a strongly coupled fixed point its conformal dimension may differ from one.  Therefore, our `anomalous' dimension is the difference of conformal dimension from the canonical value.}  For sufficiently large anomalous dimensions and range of energies in which the dynamics are approximately conformal, the operators in (\ref{operators}) may be significantly suppressed.  For this mechanism to work, the approximately conformal sector cannot have protected operators of canonical dimension that produce physical soft masses \cite{Schmaltz:2006qs}.  Conventionally, conformal sequestering has been employed in the context of anomaly-mediated supersymmetry breaking \cite{Randall:1998uk, Giudice:1998xp} to suppress additional flavor violation from gravity mediation.

However, conformal sequestering may play a similar role in gauge mediated supersymmetry breaking by lowering the scale of the soft masses.  For simplicity, consider a gauge-mediated theory involving a single set of messengers that couples the SUSY breaking sector to the MSSM.  The messengers carry gauge charges that can be embedded in representations of $SU(5)$ and have a mass $M \ll M_{Pl}$.  Below the scale $M$, the messengers may be integrated out, producing operators of the form (\ref{operators}).  If the hidden sector is approximately conformal below the scale $M,$ such operators are suppressed by the dynamics of the hidden sector.  At the supersymmetry-breaking scale $\Lambda$, conformal sequestering ends and MSSM soft masses are generated by integrating out the SUSY-breaking $F$ term $F_{S}$.  Below this scale, the contributions to renormalization from the hidden sector are negligible, and running of the soft masses may be well approximated by the MSSM RG flow.

At the scale $M$, the coefficients $c$ take the form of the ordinary gauge mediated spectrum.  The gaugino terms $c_{W}$ are generated at one loop, while the soft masses $c_{\phi}$ are generated at two loops.  We will also assume, for simplicity, that the operators that give rise to $B\mu$ and  $\mu$ operators are generated from Yukawa couplings at one loop:
\beq
\label{cmu}
c_{B \mu} \int d^4 \theta \frac{S^{\dag} S}{M^2} H_{u} H_{d} \qquad  \qquad c_{\mu} \int d^4 \theta \frac{S^{\dag}}{M} H_{u} H_{d}. 
\eeq
Finally, the operators that generate the $A$-terms arise at two loops and obtain the form $c_{A} \int d^4 \theta \frac{S}{M} \phi^{\dag} \phi$.  However, in simple models where $\mu$ is generated from a Yukawa coupling of the messeger to the Higgs, the operators $c_{A_{u,d}} \int d^4 \theta \frac{S}{M} H^{\dag}_{u,d} H_{u,d}$ are generated at one loop.  These give large contributions to the $A$ terms and can lead to significant constraints on parameter space \cite{Cho:2008fr}.\footnote{We thank Hyung Do Kim for bringing this to our attention.}

This pattern of mediation yields the following hierarchy at the scale $M:$
\beq
c_{B\mu} \gg |c_{\mu}|^2 = |c_{A_{u}} c_{A_{d}}| \simeq |c_{W}|^2 \simeq c_{\phi} \gg |c_{A}|^2.
\eeq
In the absence of strong dynamics, this spectrum gives rise to the $\mu$/$B\mu$ problem of gauge mediation \cite{Dvali:1996cu} (from this point forward we will refer to this as simply the $\mu$ problem).  Since the physics that generates $\mu$ at one loop tends to generate $B \mu$ at one loop as well, the natural prediction is $B \mu \simeq 16 \pi^2 \mu^2.$ However, satisfactory electroweak symmetry breaking requires  $B\mu \simeq \mu^2.$  In \cite{Roy:2007nz,Murayama:2007ge} it was shown that conformal sequestering offers a simple solution to this problem.

Given a sequestering hidden sector, $S$ and $S^{\dag} S$ may have large anomalous dimensions \footnote{For the sake of clarity, we are ignoring operator mixing; more accurately, the anomalous dimension properly refer to the smallest eigenvalue of a matrix of anomalous dimensions.  Alternately, one could consider a holomorphic basis where $\mu$ is not renormalized \cite{Roy:2007nz}.  In such a basis, the gravitino mass is increased by sequestering. See \cite{Roy:2007nz,Murayama:2007ge} for details.} $\gamma_{S}$ and $\gamma_{S^{\dag} S}$.  At the scale where the conformal dynamics end (taken, for simplicity, to be the SUSY-breaking scale $\Lambda$), the coefficients are given by
\beq
c_{\phi, B\mu}(\Lambda) = (\frac{\Lambda}{M})^{\gamma_{S^{\dag} S}} c_{\phi, B\mu}(M) \qquad c_{W,\mu, A}(\Lambda) =  (\frac{\Lambda}{M})^{\gamma_{S}} c_{W, \mu, A}(M).
\eeq
The MSSM soft masses then arise as $c_{W,\mu, A}(\Lambda) \frac{\Lambda^2}{M}$ or $c_{\phi, B\mu}(\Lambda) \frac{\Lambda^4}{M^2},$ using $F_{S} = \Lambda^2$.

There are several features worth highlighting.  The first, and most important for our purposes, is that all the MSSM parameters are power law suppressed compared to the na\"{i}ve mass scale of $\frac{\Lambda^2}{M}$.  Significantly, such suppression occurs relative to the gravitino mass, enabling the gravitino mass to be raised while keeping the rest of the spectrum fixed.  Secondly, provided $(\frac{\Lambda}{M})^{2 \gamma_{S}} \geq 16 \pi^2 (\frac{\Lambda}{M})^{\gamma_{S^{\dag} S}},$ the result is $B\mu \leq \mu^2$.  In other words, if the conformal dimension of $S^{\dag} S$ is more than twice that of $S$, the $\mu$ problem may be solved via sequestering.

\section{Neutralino LSP in Gauge Mediation \label{sec:theories}}

The suppression of SUSY-breaking soft masses relative to the gravitino mass raises the tantalizing possibility of obtaining a neutralino LSP from gauge mediation. In this section, we will examine the prospects for neutralino dark matter in the context of a conformally-sequestered solution to the $\mu$ problem. The pattern of mixing coefficients arising from this solution to the $\mu$ problem leads to distinctive boundary conditions at the scale of supersymmetry breaking. At the weak scale, this results in a characteristic sparticle spectrum admitting neutralino dark matter with satisfactory relic abundance.

In order to obtain a non-gravitino LSP, it must be the case that the gravitino mass is larger than the other soft masses at the weak scale. Given $m_{\frac{3}{2}} \sim \Lambda^2 M_{Pl}^{-1},$ this suggests a high gauge-mediated supersymmetry-breaking scale $\Lambda \gtrsim 10^{10-11}$ GeV. Since there is a gravity mediation contribution to $B\mu$ proportional to $\mu m_{\frac{3}{2}},$ preserving the sequestered solution to the $\mu$ problem likewise indicates $\Lambda \sim 10^{10-11}$ GeV.

\begin{figure}[t] 
	   \centering
	   \includegraphics[width=2in]{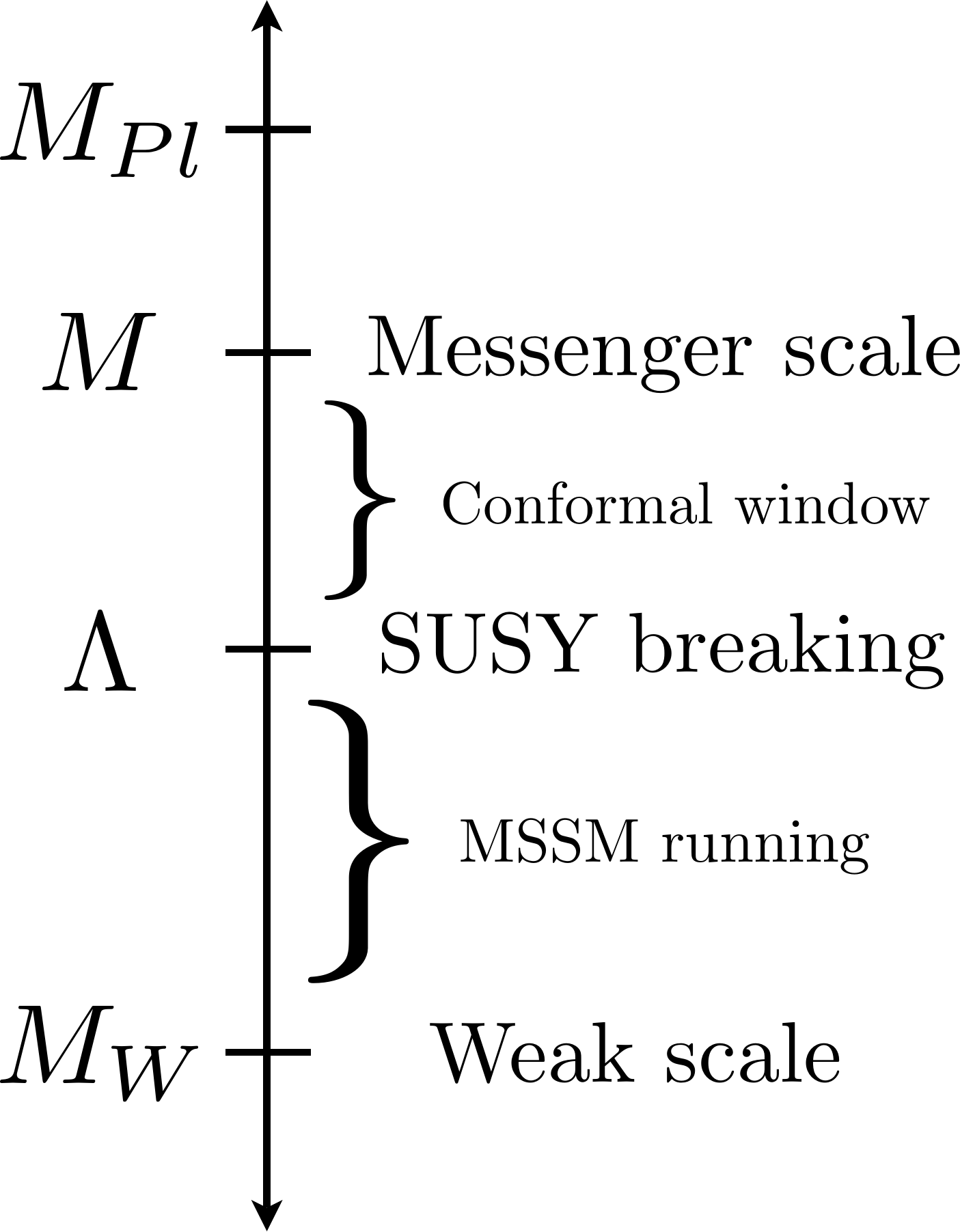} 
	   \caption{ Scales and dynamics for conformally-sequestered gauge mediation with neutralino LSP. Here $M \simeq 10^{13-14}$ GeV and $\Lambda \simeq 10^{10-11}$ GeV.}
	   \label{fig:0}
	\end{figure}

In a particular conformal sector, the anomalous dimensions are strictly determined. In the interest of generality, however, we will refrain from specifying the exact conformal dynamics, and instead consider a reasonably generic range of parameters. We will assume that $\gamma_{S}  \sim \mathcal{O}(1) $ and $\gamma_{S^{\dag} S} - 2 \gamma_{S} \sim \mathcal{O}(1) > 0$.  This is often the case in strongly coupled theories with known gravity duals \cite{Ceresole:1999ht, Aharony:2005ez}.  These assumptions are made to ensure that sequestering produces an adequate hierarchy of scales and that higher dimensions operators are unimportant.  Given these assumptions, we can relate the messenger scale to the anomalous dimension by requiring that the soft masses are at the weak scale $\sim 100$ GeV.  Specifically, we require that
\beq
\frac{g^2}{16\pi^2}\Lambda (\frac{\Lambda}{M})^{1+\gamma_{S}} \sim m_{Z},
\eeq
where the factor of $\frac{g^2}{16\pi^2}$ comes from the origin of the coefficient $c_{W}$ via a one-loop diagram.  For $\gamma_{S} > 1$ we obtain $M < 10^{13-14}$ GeV.  Under this assumption, the gravity-mediated contributions to the soft masses are suppressed by $\frac{M}{M_{Pl}} < 10^{-5}$.  Therefore, with reasonable anomalous dimensions, the gravitino may be made heavier than the bino without reintroducing the flavor problem. The corresponding hierarchy of scales is illustrated in Fig. \ref{fig:0}.

If we retain a solution to the $\mu$ problem, the quadratic operators involving $S^\dagger S$ must be suppressed by at least $(16\pi^2)^{-1}$ relative to the operators linear in $S.$  At the messenger scale $M$, ordinary gauge mediation predicts that the scalar and gaugino masses will be approximately equal.  Therefore, at the SUSY-breaking scale $\Lambda$, we should have $m_{gaugino}^2(\Lambda) \geq 16 \pi^2 m_{scalar}^2 (\Lambda)$.

Below the scale $\Lambda$, the running of soft masses may be well approximated by the MSSM renormalization group.  In light of our previous discussion, the gaugino masses ($M_{i}$), $\mu$ and $B \mu$ are near the weak scale.  For simplicity, we take the gaugino masses to satisfy the usual GUT relations,  $M_{i} \alpha_{i}^{-1} = M_{j} \alpha_{j}^{-1}$.  The squark and slepton masses will be suppressed relative to the gaugino masses by at least of factor of $4 \pi$.  In the case where $(\frac{\Lambda}{M})^{2 \gamma_{S}} \gg16 \pi^2 (\frac{\Lambda}{M})^{\gamma_{S^{\dag} S}},$, then it is consistent to take $\tilde{m}_{squark} \simeq \tilde{m}_{slepton} \simeq 0$.  There are additional contributions to the scalar masses from $c_{A}$, but these arise at two loops.  Therefore we will ignore these contributions, as they do not significantly change our results.

There is a subtlety that arises with $B\mu$ and the Higgs soft masses $m_{H_{u}}^2$ and $m_{H_{d}}^2$.  As explained in \cite{Roy:2007nz}, the operators that generate $A_{u,d}$ and $\mu$ in (\ref{cmu}) also contribute to the running of couplings $c_{B\mu}$, $c_{H_{u}}$ and $c_{H_{d}}$ during sequestering.  For example, for $c_{H_{u,d}}$ this produces an additional contribution to the beta function proportional to $|c_{\mu}|^2$.  We expect that this will also depend on a factor coming from the CFT operator product expansion
\beq
S^{\dag}(x) S (y) \sim C |x-y|^{(\gamma_{S^{\dag} S}-2 \gamma_{S})} S^{\dag} S (y)
\eeq
Therefore, the beta function for $c_{H}$ takes the form $\beta(c_{H_{u,d}}) \sim \gamma_{S^{\dag} S} c_{H_{u,d}} + 2 \pi^2 C (|c_{\mu}|^2+|c_{A_{u,d}}|^2)$.   For $B \mu$, the beta function takes a similar form $\beta(c_{B \mu}) \sim \gamma_{S^{\dag} S} c_{B \mu} + 2 \pi^2 C c_{\mu} (c_{A_{u}} + c_{A_{d}})$.  Thus at the scale $\Lambda$, we have $|m_{H_{u,d}}^2| \sim |\mu^2|$ and $B \mu \sim \mu^2$, although the precise values will depend on details of the conformal dynamics.\footnote{Here we disagree with the claims of \cite{Murayama:2007ge}, that argue $m_{H_{u,d}}^2 =-\mu^2$.  For more discussion of the hidden sector contribution to the running see \cite{Cohen:2006qc, Roy:2007nz, Craig}}

Combining the above constraints, we obtain the following high scale boundary conditions:
\beq
M_{i} = g_{i}^{2} M_{0} \qquad \tilde{m}_{q} \sim \tilde{m}_{L} \sim 0 \qquad m_{H_{u,d}}^2 \sim \mu^2 \sim A_{u,d}^2 \sim B\mu_{grav} \sim B\mu_{gauge} \sim M_{i} \gg A^2,
\label{bcs}
\eeq
where $M_{i}$ are the gaugino masses ($i=1,2,3$ refers to the $U(1)$, $SU(2)$ or $SU(3)$ gauge group) and $\tilde{m}_{q}$ and $\tilde{m}_{L}$ are the (universal) squark and slepton masses, respectively.  $B\mu_{gauge}$ is the contribution to $B\mu$ from the operator in (\ref{cmu}) and $B\mu_{grav} = m_{\frac{3}{2}} \mu$ is the contribution to $B\mu$ from gravity/anomaly mediation (which is unsequestered).  This spectrum bears a close resemblance to that of gaugino mediation \cite{Kaplan:1999ac, Chacko:1999mi}.  In the case where we arrange just enough sequestering to have $B\mu_{gauge} \simeq \mu^2$ then we would have $\tilde{m}_{q,L}^2(\Lambda) = (16\pi^2)^{-1} \tilde{m}_{q,L}^2(M)$.   $M_{0}$ and the parameters of the higgs sector are variables that will be determined by requiring satisfactory electroweak symmetry breaking and neutralino LSP.

Given these conditions at the scale $\Lambda$, we run them to the low scale $M_S = \sqrt{m_{\tilde t_1}(M_S) m_{\tilde t_2}(M_S)}$ (at which point the scale dependence of electroweak breaking conditions is smallest) using SoftSUSY \cite{Allanach:2001kg}. The basic features of flow are easy to understand.  The gaugino masses unify at the GUT scale and run with their couplings.  Therefore, the lightest gaugino will be the bino ($M_{1}$) because it is associated with the $U(1)$ of the MSSM.  The squarks and sleptons will run positive due to the contribution from gaugino masses.  The right-handed sleptons are only charged under the $U(1),$ and thus they will run more slowly than the squarks and other sleptons.  Of these, the stau is the often the lightest because it has the largest Yukawa, which enters the beta function with opposite sign. Although the sneutrino accumulates a larger soft term at the low scale, Standard Model D-term contributions to the physical sneutrino mass make it competitive with the stau. Therefore, depending on the choice of high-scale parameters, the LSP tends to be a neutralino, stau, or sneutrino.  

The stau is charged, and hence is not a good dark matter candidate. In the MSSM, the left-handed sneutrino is likewise a poor DM candidate; rapid annihilations via Z-mediated s-channel diagrams yield insufficient relic abundance, and attempts to suppress annihilation rates with low or high sneutrino mass have been ruled out by precision electroweak and direct detection experiments, respectively. The bino, however (which becomes a component of the lightest neutralino after electroweak symmetry breaking), may be an excellent candidate for cold dark matter.  We are interested in the range of parameters where the bino is the lightest.  As the stau, sneutrino, and bino masses are all roughly degenerate at the low scale, the LSP in this scenario depends rather sensitively on high scale parameters. In the next section, we will consider the details of the weak-scale spectrum and parameter space for neutralino dark matter.

\section{Parameter Space for Neutralino Dark Matter \label{sec:dm}}

In the previous section, we have examined the high scale boundary conditions arising from sequestered gauge mediation with an eye towards non-gravitino LSP. For this LSP to be a satisfactory dark matter candidate, it must be stable; this is easily accomplished using R-parity, which we will assume is a good symmetry.  For a cold relic, the relic abundance can be calculated in general and is given by $\Omega \propto \langle \sigma v \rangle^{-1}$.  Given a dimensionless coupling $g$ and a mass $m$, the cross section is roughly $\langle \sigma v \rangle \sim g^4 m^{-2}$.  For a weakly interacting particle with a mass of 100 GeV to 1 TeV, this gives about the right relic abundance, leading to the common observation that dark matter independently predicts new physics at the weak scale.

A weakly interacting particle in this mass range is no guarantee that the relic abundance will be consistent with observation; in the case of neutralinos, the details may be difficult to arrange. Specifically, the annihilation amplitude is helicity suppressed, making the cross section smaller by a factor of eight.  There are many ways of getting around this problem in different regions of parameter space. One well-known regime where the relic abundance works out correctly is when the mass of the relic is taken to be near or below $100$ GeV, simply because the lower mass increases the cross section to an acceptable level.  It is a complete coincidence that our model naturally suggests this low mass range for dark matter.

In addition to these general constraints for acceptable dark matter, there arise specific experimental constraints from direct detection and collider experiments that strongly constrain the parameter space of our model. Although flavor invariance of SUSY-breaking soft terms is one of the principal theoretical successes of gauge mediation, constraints from FCNCs may still arise from processes sensitive to new particle exchange, such as the inclusive B-meson decay $B \rightarrow X_s \gamma.$ For moderate values of $\tan \beta$ such as those considered here, the dominant contribution arises from top quark and charged Higgs loops; these processes interfere constructively with Standard Model contributions. The 95\% CL upper limit on $BR(B \rightarrow X_s \gamma)$ from CLEO \cite{Alam:1994aw} leads to the limit $m_{H^\pm} > 340$ GeV \cite{Ciuchini:1997xe}. 

	The characteristic spectrum of light sleptons arising from conformally-sequestered gauge mediation suggests that experimental bounds from LEP2 may play a significant role in constraining the dark matter parameter space. Although precise bounds are quite model-dependent, we require the low energy spectrum to satisfy the general sparticle limits from non-observation at LEP2 \cite{Barate:2003sz, Alcaraz:2007ri}.
	
Although realistic electroweak symmetry breaking is not our principal concern, we are likewise interested in satisfying rudimentary constraints on the stability of the Higgs scalar potential. Requiring that the Higgs scalar potential possesses nontrivial extrema (i.e., $\langle h_u^0 \rangle, \langle h_d^0 \rangle \neq 0$) entails $(B \mu)^2 > (m_{H_u}^2 + \mu^2) (m_{H_d}^2 + \mu^2),$ while ensuring that the scalar potential possesses a stable minimum leads to $m_{H_u}^2 + m_{H_d}^2 + 2 \mu^2 > 2 | B\mu |.$ We likewise take account of direct search bounds on the lightest neutral Higgs mass, $m_h > 114$ GeV, allowing for $\pm 3$ GeV due to theoretical errors among spectrum-calculating software packages.

	We have used the software package MicrOMEGAs \cite{Belanger:2001fz, Belanger:2004yn, Belanger:2006is, Belanger:2008sj} to compute the dark matter relic abundance and detection rates from the low-scale soft parameters. Naturally, the high-scale boundary conditions (\ref{bcs}) admit a multidimensional parameter space. The values of $\mu, \, B, A_{u,d},$ and the gaugino masses $M_i = g_i^2 M_0$ are set at the high scale, while the electroweak parameter $\tan \beta \equiv \frac{v_u}{v_d}$ is chosen at the low scale. Large values of $\tan \beta$ raise the $\tau$ yukawa coupling, further lowering slepton masses at the weak scale; this leads us to favor small values of $\tan \beta$ consistent with LEP bounds on Higgs and sparticle masses. We then examine dark matter candidates as a function of $\mu$ and the universal gaugino mass $M_0.$ The choices of $\mu, B,$ and $\tan \beta$ fix the high-scale values of $m_{H_u}^2$ and $m_{H_d}^2$ in order to satisfy the conditions for electroweak symmetry-breaking.  

\begin{figure}[t] 
	   \centering
	   \includegraphics[width=3in]{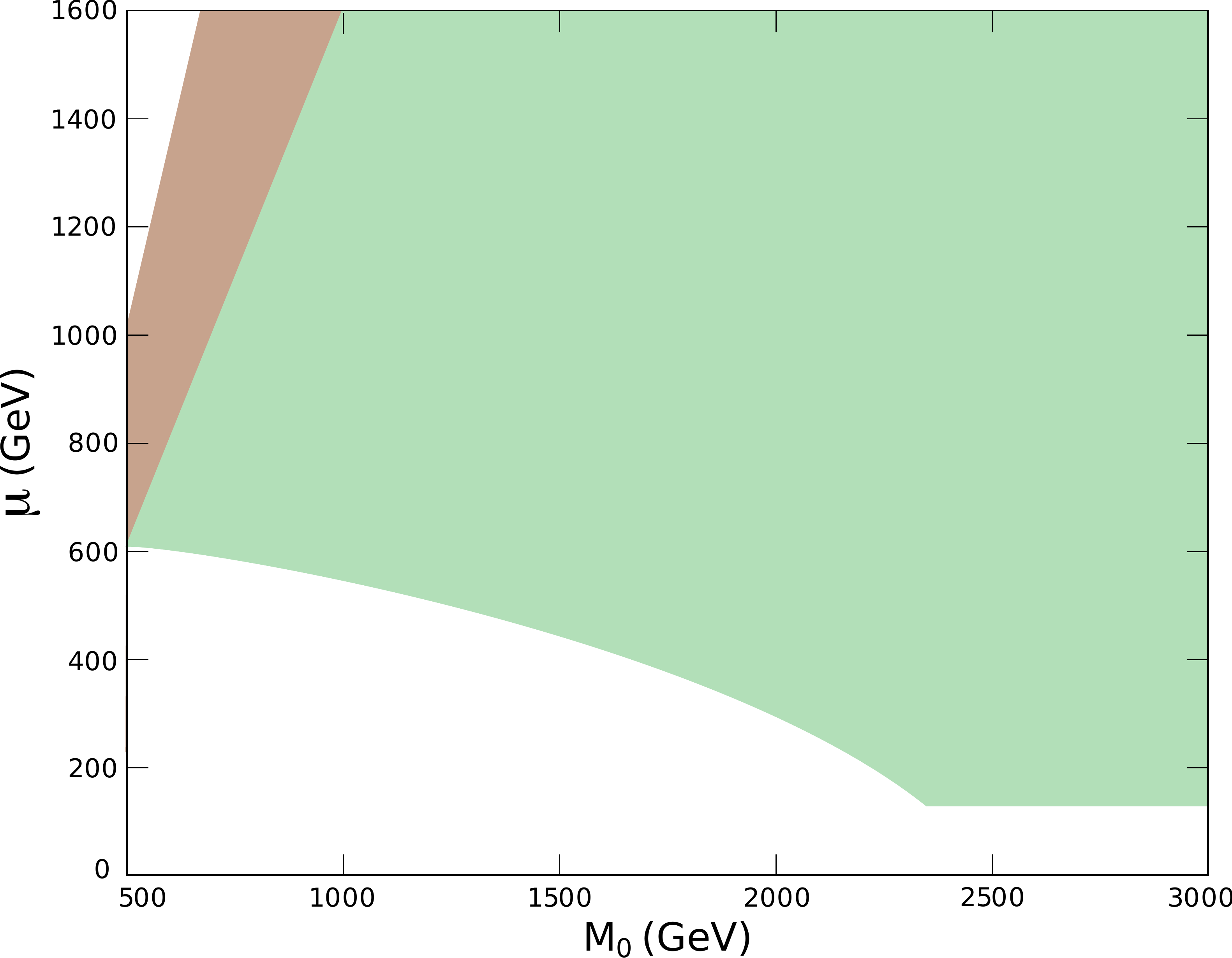} 
	   \caption{Lightest supersymmetric particle as a function of $\mu$ and universal gaugino mass parameter $M_0$ for the case of maximal sequestering ($\tilde m_{q}(\Lambda) = \tilde m_{L}(\Lambda) = 0$) with $\tan \beta = 5,  \, \Lambda = 10^{11} \text{ GeV}, \, A_{u,d} = \mu.$ Green regions denote stau dark matter, while brown regions denote sneutrino dark matter. Regions excluded by LEP2 sparticle mass limits and $B \rightarrow X_s \gamma$ are shown in white.}
	   \label{fig:a}
	\end{figure}

	The high-scale boundary conditions (\ref{bcs}) yield a highly degenerate parameter space in which the stau, sneutrinos, and lightest neutralino obtain physical masses of order $\sim 100$ GeV. In order to fully elucidate the parameter space of dark matter candidates, we consider two cases of conformal sequestering. The first case is maximal suppression of the gauge-mediated contributions to $B \mu,$ in which $B \mu_{gauge} \sim 0$ at the supersymmetry-breaking scale $\Lambda;$ this leads to complete suppression of high-scale squark and slepton masses, $\tilde m_{q}(\Lambda) \sim \tilde m_{L}(\Lambda) \simeq 0.$ The resultant low-scale parameters favor a light stau and sneutrinos, which tends to produce unacceptable dark matter candidates for wide regions of $\mu$ and $M_0.$ As illustrated in Fig. \ref{fig:a}, although the bino is light, the LSP is stau or sneutrino for the entire range of $\mu$ and $M_0;$ maximal sequestration seems maximally unsuited for neutralino dark matter.
	
\begin{figure}[t] 
	   \centering
	   \includegraphics[width=6in]{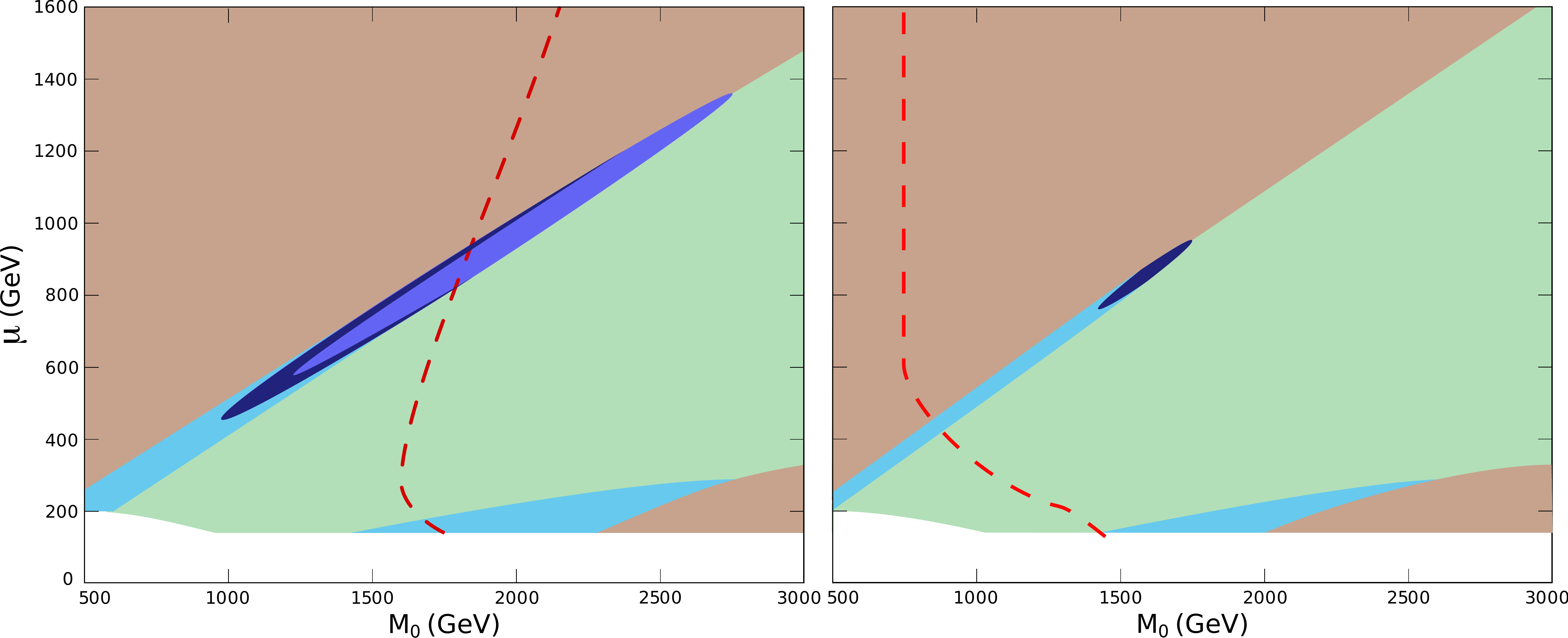} 
	   \caption{Lightest supersymmetric particle as a function of $\mu$ and universal gaugino mass parameter $M_0$ for the case of minimal sequestering ($\tilde m_{q}(\Lambda) = \tilde m_{L}(\Lambda) = M_0 / 4 \pi$) with $\tan \beta = 5, \, \Lambda = 10^{11} \text{ GeV}.$ On the left, $A_{u,d} = \mu;$ on the right, $A_{u,d} = - \mu.$  Regions of neutralino dark matter are blue; regions of stau dark matter are green; regions of sneutrino dark matter are brown. In the light blue region, neutralino relic abundance is below the WMAP lower bound $\Omega_{DM} h^2 < 0.094,$ while in the dark blue region the neutralino relic abundance lies within the WMAP bounds $0.094 < \Omega_{DM} h^2 < 0.129.$ In the medium blue region, relic abundance lies within pre-WMAP bounds $0.129 < \Omega_{DM} h^2 < 0.3$. The neutral Higgs mass is too light to the left of the red dotted line, which denotes $m_h \geq 111.4$ GeV. Regions excluded by LEP2 sparticle mass limits and $B \rightarrow X_s \gamma$ are shown in white.}
	   \label{fig:b}
	\end{figure}

	The second case is minimal suppression of the gauge-mediated contributions to $B \mu,$ in which we assume only enough conformal sequestering to solve the $\mu$ problem and yield $B \mu_{gauge} \simeq \mu^2.$ This, in turn, leads to small but nonzero squark and slepton masses at the high scale, $\tilde m_{q}^2(\Lambda) \sim \tilde m_{L}^2(\Lambda) \simeq M_0^2/16 \pi^2.$ The degeneracy of stau, sneutrino, and bino masses is such that even this small contribution to slepton masses at the high scale significantly expands the range of neutralino LSP and satisfactory relic abundance, as seen in Fig. \ref{fig:b}. For a wide range of $\mu$ and $M_0,$ the LSP is a mostly-bino neutralino with relic abundance $\Omega_{DM} h^2 = 0.1 - 0.3.$ Although the relic abundance of bino-like neutralino dark matter is often prohibitively high, the near-degeneracy of the stau, sneutrino, and neutralino lowers the relic abundance through coannihilations.  In particular, there exists a region with mostly-bino neutralino LSP satisfying both WMAP bounds on relic abundance, $0.094 < \Omega_{DM} h^2 < 0.129,$ and LEP2 bounds on sparticle and Higgs masses for either sign of $A_{u,d}.$ For small values of $\mu$ relative to the bino mass $M_1$ there also arises a region of mostly-higgsino neutralino, although here the relic abundance is well below the WMAP lower bound.
	
These two cases are simply intended to be illustrative; there naturally exists a continuum of spectra interpolating between minimal and maximal conformal sequestering, with correspondingly varying regions of parameter space exhibiting neutralino LSP and suitable relic abundance. Due to the near-degeneracy of bino, stau, and sneutrino, the region of parameter space with neutralino LSP decreases with increasing amounts of conformal sequestering. Nonetheless, neutralino LSP in conformally-sequestered gauge mediation appears to be generic for a wide range parameters. 
	
	One should keep in mind that the specific mass spectra resulting from this model arise under many unnecessary made strictly for the sake of clarity. Specifically, we have used sequestering to solve the $\mu$ problem, which forces the rest of the masses into a fairly limited region of parameter space.  Furthermore, we have assumed a spectrum at the scale $M$ consistent with a single-messenger model.  Model building in the messenger sector and the SUSY breaking sector can easily change the spectrum without reintroducing the $\mu$ problem.  For example, additional D-term contributions can be used to increase the mass of the scalars, while relaxing GUT-scale gaugino mass unification would allow for a relatively lighter bino. Alternatively, models with R-symmetry breaking at a low scale could also balance the hierarchy between the gaugino and scalar masses.   Allowing these additional effects or using the messenger sector to solve the  $\mu$ problem, one could then construct many models with neutralino LSP/dark matter where the parameter space is larger and the mass is much higher than we have discussed here.  The expanded parameter space admitted by more general models will be discussed in a subsequent publication \cite{Craig}.

\section{Conclusion}

In this letter, we have demonstrated that neutralino dark matter is possible in gauge mediation.  Specifically, the $\mu$ problem and flavor problem may be simultaneously solved in a simple model of gauge mediation with conformal sequestering, in which a bino-like neutralino is the lightest supersymmetric particle.  Moreover, the relic abundance of this neutralino dark matter may be consistent with cosmological observations.

In the coming years, the LHC will give us a better picture of physics beyond the Standard Model.  Likewise, dark matter direct detection experiments will increasingly probe some of the most interesting ranges of parameter space. Here we provide an example of a very general class of theories producing satisfactory signatures in both theaters: dark matter within range of direct detection experiments and a characteristic sparticle spectrum that is both consistent with current collider bounds and within reach of future experiments.

While this work was being completed, \cite{Shirai:2008qt} appeared which uses a similar mechanism to obtain neutralino dark matter in gauge mediation.  Their model employs both a separate solution to the $\mu$ problem and sequestering above the messenger scale to solve the flavor problem.

\acknowledgments
We would like to thank Xiao Liu for collaboration at various stages of this project.  We would also like to thank Savas Dimopoulos, Michael Dine, Shamit Kachru, Hyung Do Kim, Michael Peskin, Eva Silverstein, and especially David Poland for helpful discussions.  NJC would like to acknowledge the hospitality of the Rudolph Peierls Center for Theoretical Physics at Oxford University, where part of this work was completed. DG is supported in part by NSERC, the Mellam Family Foundation, the DOE under contract DE-AC03-76SF00515 and the NSF under contract 9870115. NJC is supported in part by the NDSEG Fellowship, the NSF under contract PHY-9870115, and the Stanford Institute for Theoretical Physics.

\bibliography{gmdm}
\end{document}